\documentclass[preprint,12pt,showkeys,nofootinbib,floatfix]{revtex4}

\usepackage{fancyhdr}
\usepackage{amssymb}
\usepackage{amsfonts}
\usepackage{amsmath}
\usepackage{graphicx}
\usepackage{subfigure}
\usepackage{tikz}
\usepackage[
            pdfstartview=FitH,
            bookmarksnumbered=true,
            bookmarksopen=true,
            colorlinks,
            pdfborder=001,
            linkcolor=green,
            anchorcolor=green,
            citecolor=red
            ]{hyperref}
\usepackage[toc,page,title,titletoc,header]{appendix}
\usepackage{graphics}
\usepackage{epsfig}
\usepackage{slashed}
\usepackage{latexsym}
\usepackage{syntonly}
\newcommand{\be}{\begin{equation}}
\newcommand{\ee}{\end{equation}}
\newcommand{\bea}{\begin{eqnarray}}
\newcommand{\eea}{\end{eqnarray}}

\begin{document}
\title{Holography as deep learning}
\author{Wen-Cong Gan$^{1,2}$}
\author{Fu-Wen Shu$^{1,2}$}
\thanks{Corresponding author}
\thanks{E-mail address:shufuwen@ncu.edu.cn}
\affiliation{
$^{1}$Department of Physics, Nanchang University, No. 999 Xue Fu Avenue, Nanchang, 330031, China\\
$^{2}$Center for Relativistic Astrophysics and High Energy Physics, Nanchang University, No. 999 Xue Fu Avenue, Nanchang 330031, China}
\begin{abstract}
Quantum many-body problem with exponentially large degrees of freedom can be reduced to a tractable computational form by neural network method \cite{CT}. The power of deep neural network (DNN) based on deep learning is clarified by mapping it to renormalization group (RG), which may shed lights on holographic principle by identifying a sequence of RG transformations to the AdS geometry. In this essay, we show that any network which reflects RG process has intrinsic
hyperbolic geometry, and discuss the structure of entanglement encoded in the graph of DNN. We find the entanglement structure of deep neural network is of Ryu-Takayanagi form. Based on these facts, we argue that the emergence of holographic gravitational theory is related to deep learning process of the quantum field theory.
\end{abstract}
\maketitle
\thispagestyle{fancy}        
\fancyhead{}                     
\lhead{Essay written for the Gravity Research Foundation 2017 Awards for Essays on Gravitation.}      
\chead{}
\rhead{}
\lfoot{}
\cfoot{\thepage}   
\rfoot{Submission date: March 31}
\renewcommand{\headrulewidth}{0pt}       
\renewcommand{\footrulewidth}{0pt}

\pagestyle{plain}
\cfoot{\thepage}
\section{Introduction}
Solving quantum many-body problem with exponentially large degrees of freedom is a central task in modern condensed matter field theory. The key point is to find efficient representation in which quantum many-body state can be reduced to a tractable computational form.
Neural network is a powerful artificial intelligence technique in representing complex correlation and finds its applications in representing quantum many-body state recently \cite{CT}.
The deep reason why neural network can easily extract relevant features from complex data had been clarified by mapping it to variational renormalization group \cite{MS} which plays a key role in understanding features in complex quantum many-body problems.

Another modern tool to tackle quantum many-body problem is based on AdS/CFT correspondence\cite{maldacena1,gkp,witten} which relates a quantum field theory to a gravitational theory on anti de-Sitter space with one higher dimension and thus is a realization of holographic principle \cite{hooft,susskind}. In AdS/CFT, a sequence of renormalization group (RG) transformations of the CFT on the boundary of AdS are identified to the AdS geometry \cite{JG}. In this essay, we argue that the emergence of holographic gravitational theory is intimately related to the deep learning process of the partition function of a quantum field theory.

\section{Renormalization group as deep learning}
\begin{figure}[t]
\centering 
\includegraphics[width=.55\textwidth]{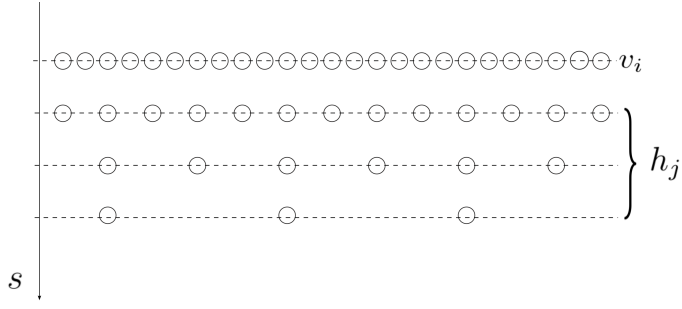}

\caption{\label{fig:0} Renormalization process of one dimensional system. In the figure, every two adjacent spins are coarse grained to one effective spin in next layer.}
\end{figure}

In this section, we briefly review the equivalence between renormalization group and deep learning based on \cite{MS}.
We study $N$ binary spins \{$v_i$\} ($i=1,2 \cdots N$) in Boltzmann distribution
\begin{equation}
P(\{v_i\})=\frac{e^{-\mathbf H(\{v_i\})}}{Z}=\frac{e^{ \sum_i K_i v_i +\sum_{ij} K_{ij} v_i v_j + \sum_{ijk} K_{ijk} v_i v_j v_k +\cdots}}{Z},
\end{equation}
where $H(\{v_i\})$ is the Hamiltonian, and $Z$ is the partition function
\begin{equation}
Z=\mathrm Tr_{v_i}e^{-\mathbf H(\{v_i\})}\equiv \sum_{v_1,\cdots v_N=\pm 1}e^{-\mathbf H(\{v_i\})}.
\end{equation}
where $\mathbf K=\{K_s  \}$ is coupling constants. After renormalization, coarse-grained spins $\{h_j\}$ are introduced and the effective Hamiltonian becomes
\begin{equation}
\mathbf H^{RG}(\{h_j\})=-\sum_i \tilde K_i h_i -\sum_{ij} \tilde K_{ij} h_i h_j - \sum_{ijk} \tilde K_{ijk} h_i h_j h_k +\cdots,
\end{equation}
where the coupling constants is renormalized to be $ \tilde{\mathbf K}=\{\tilde K_s  \}$. This is one step of renormalization. Repeating the above step, more effective spins are introduced, and all of the steps consist the renormalization group as depicted in Fig.\ref{fig:0}, where each layer stands for one step of the RG.

In the RG scheme proposed by Kadanoff, RG process can be understood as follows
\begin{equation}
e^{-\mathbf H^{RG}(\{h_j\})}\equiv \mathrm Tr_{v_i}e^{\mathbf T_\lambda (\{v_i\},\{h_j\})-\mathbf H(\{v_i\})},
\end{equation}
where $\mathbf T_\lambda (\{v_i\},\{h_j\})$ is an adjustable function with parameters $\{\lambda \}$ which should be tuned to make the RG transformation approximately exact.

To model data distribution, neural network also introduces new spin variables $\{h_j\}$ which is called \emph{hidden neurons} in the language of neural network. In contrast, physical spin variables $\{v_i\}$ are called \emph{visible neurons}. Hidden neurons and visible neurons compose the restricted Boltzmann machine (RBM). Thus the graph of RBM has the same structure of RG in Fig.\ref{fig:0} and step of RG is called \emph{depth} in deep learning. The probability of a configuration of hidden neurons is given by
\begin{equation}
p_\lambda(\{h_j\})= \sum_{\{v_i\}}\frac{e^{-\mathbf E(\{v_i\},\{h_j\})}}{\mathcal Z}=\sum_{\{v_i\}}\frac{e^{-\sum_j b_j h_j - \sum_{ij} v_i w_{ij} h_j -\sum_i c_i v_i}}{\mathcal Z},
\end{equation}
where $\lambda=\{b_j, w_{ij}, c_i\}$ are adjustable parameters, and the RBM Hamiltonian for hidden neurons is defined as
\begin{equation}
p_\lambda(\{h_j\})= \frac{e^{-\mathbf H^{RBM}_{\lambda}(\{h_j\})}}{\mathcal Z}.
\end{equation}
Above analysis implies there is a following mapping between renormalization group and deep learning
\begin{align}
\mathbf T_\lambda (\{v_i\},\{h_j\})&=-\mathbf E(\{v_i\},\{h_j\})+\mathbf H(\{v_i\}), \\
\mathbf H^{RG}_{\lambda}(\{h_j\})&=\mathbf H^{RBM}_{\lambda}(\{h_j\}).
\end{align}
Therefore deep neural network (DNN) can be exactly mapped to RG.

\section{Emergent spacetime from renormalization group}
In this section,we will show that any network which reflects RG process has intrinsic hyperbolic geometry.

\begin{figure}[h]
\centering 
\includegraphics[width=.55\textwidth]{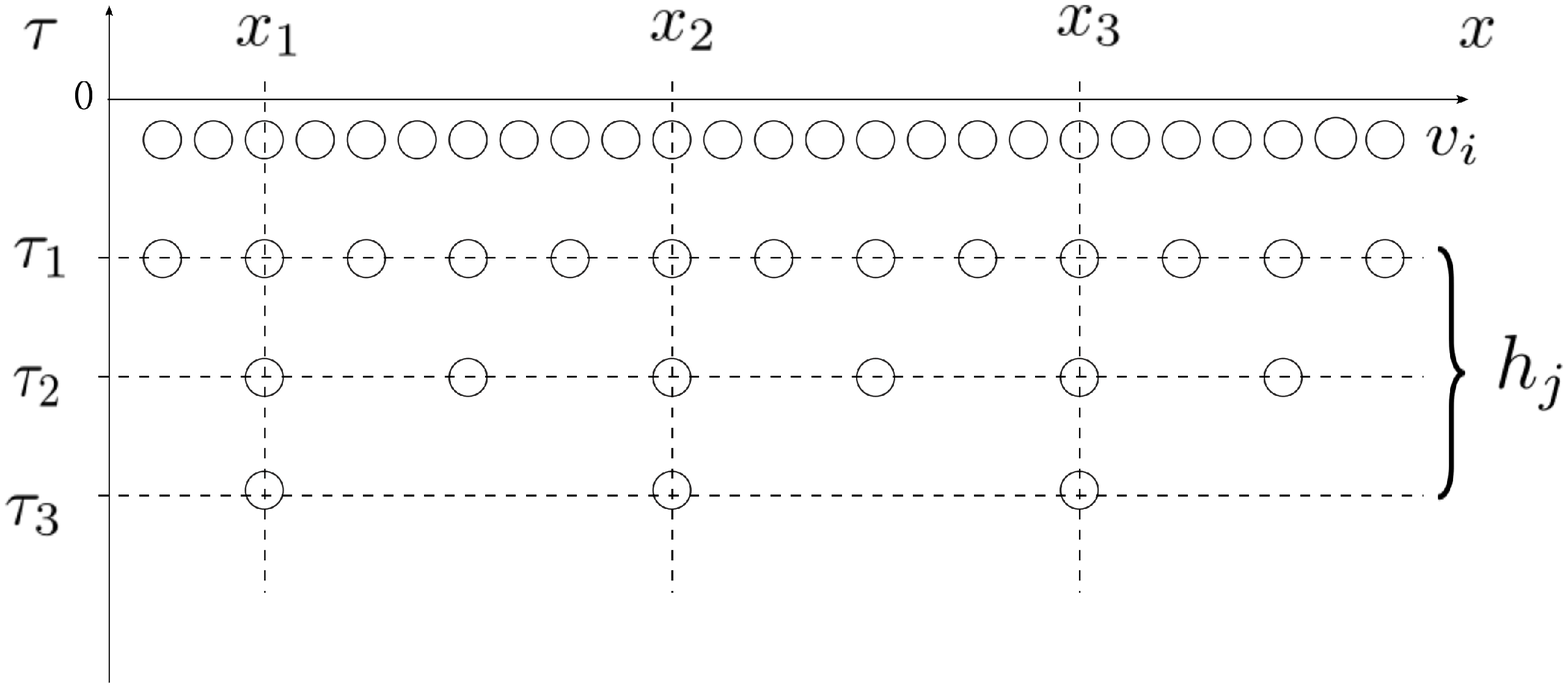}

\caption{\label{fig:1} Renormalization process of one dimensional system, where every two adjacent spins are coarse grained to one effective spin in next layer.}
\end{figure}

In Fig.\ref{fig:1}, we consider one dimensional spin system with translational invariance. Nodes stand for spin variables, the first line is physical spin $\{v_i\}$, and remains are coarse-grained spins $\{h_j\}$. The RG process is from up to down. We define the coordinates $(\tau,x)$ as follows: every coarse-grained spin is in the same equal-$x$ line, and spins in every step of RG are in the equal-$\tau$ line as depicted in Fig.\ref{fig:1}. Then the metric ansatz for this graph can be written as
\begin{equation}
ds^2=f_1(\tau) d\tau^2 + f_2(\tau) dx^2
\end{equation}
where $f_{1,2}$ does not depend on $x$ because of the translational invariance. We assume that the proper volume of some region is proportional to the number of nodes in the region. Then the proper distance between $x_1$ line and $x_2$ line is given by
\begin{equation}
\int_{x_1}^{x_2} \sqrt{f_2(\tau)} dx = \sqrt{f_2(\tau)} \Delta x = a 2^{-\tau},
\end{equation}
where $a$ is the proper distance between $x_1$ and $x_2$ when $\tau=0$. Then
\begin{equation}
f_2(\tau) = \alpha e^{-2\tau},
\end{equation}
where $\alpha$ is some constant.
The proper distance between $\tau_1$ and $\tau_2$ should be a constant because we place every layer of spins on the same $\tau$.
\begin{equation}
\int_{\tau_1}^{\tau_2} \sqrt{f_1(\tau)} d\tau = constant,
\end{equation}
then $f_1(\tau)=constant$. This is a free parameter and we can safely set it to $\alpha$. Then the metric becomes
\begin{equation}
ds^2=\alpha[ d\tau^2 + e^{-2\tau} dx^2],
\end{equation}
which is recognized as Euclidean dS$_2$, and after coordinate transformation $\tau=\log z$, it cames to
\begin{equation}
ds^2=\frac{\alpha}{z^2}[ dz^2 + d x^2],
\end{equation}
which is the metric of hyperbolic plane H$_2$.

We would like to comment that scale transformation consists the coarse grain process and it is isometric, the only Euclidean geometry which is invariant under isometry of scale transformation is Euclidean dS$_d$ and H$_d$.

\section{Entanglement entropy from deep neural network}
The deep neural network(DNN) or RBM of renormalization process can be depicted as \cite{MS} where only nearby nodes are linked, and we call it short-ranged RBM. In \cite{DLS}, the authors proved that all short-ranged RBM states exhibit \emph{area-law} entanglement. In this section, we will show how the structure of entanglement is encoded on the graph of deep neural network.

\begin{figure}[h]
\centering 
\includegraphics[width=.55\textwidth]{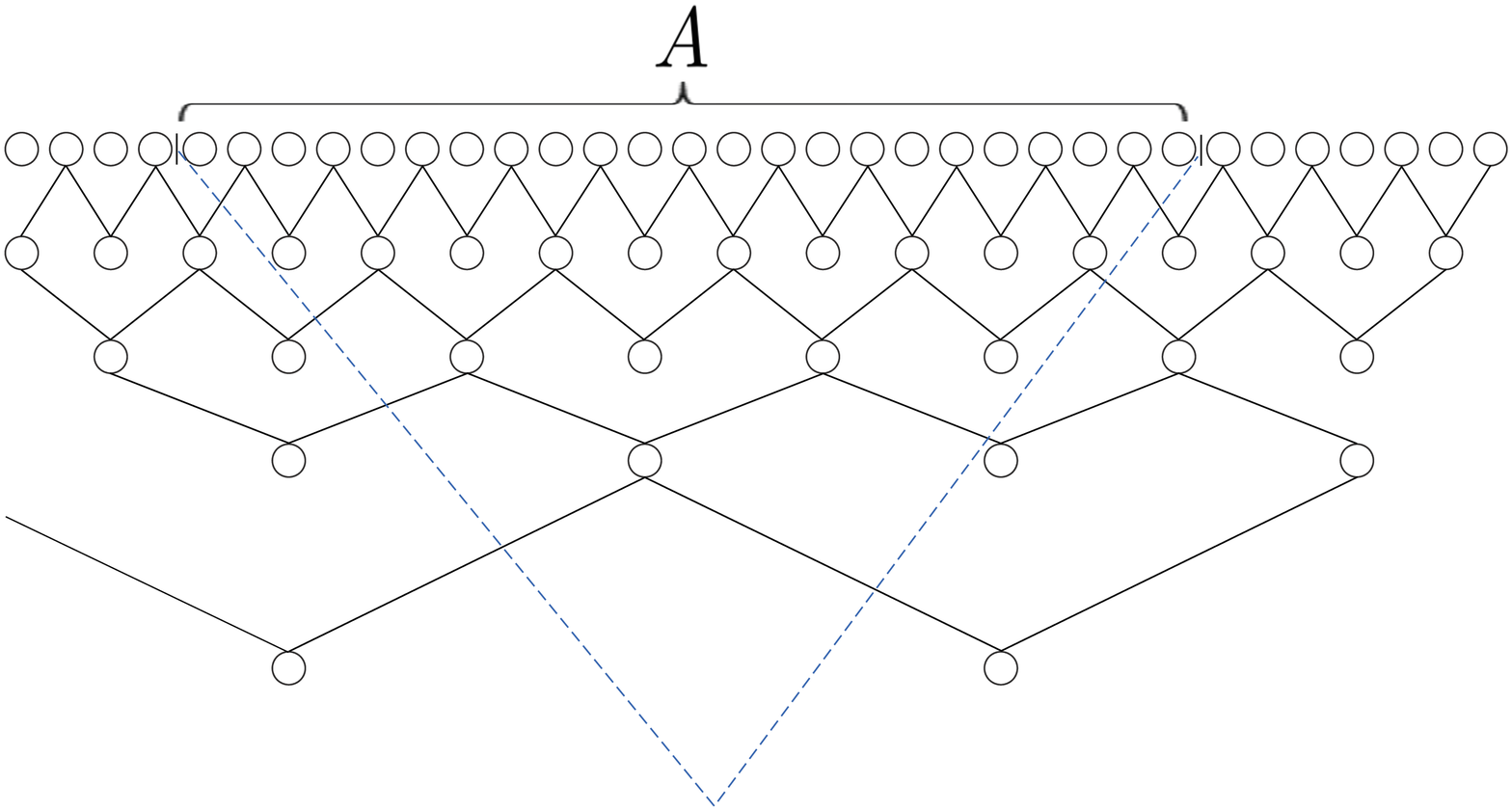}

\caption{\label{fig:2} Deep neural network representation of renormalization of one dimensional spin system. The link between neurons stands for $w_{ij}$.}
\end{figure}

We choose a sub-region $A$ consisting $\tilde N$ nodes of the physical spin system. After $\tau= \log_2 \tilde N$ steps of RG, $A$ is coarse-grained to one node. In Fig.\ref{fig:2} we can see that the minimal number of links that should be cut to apart $A$ from the remained region is given by $n \propto\tau$, because on average across every layer there is one link should be cut, as depicted by dashed line in Fig.\ref{fig:2}. As pointed in \cite{HM} every local tensor network has neural network representation, we argue that the entanglement entropy $S_A$ of $A$ is proportional to the number of links that are cut just like the case of tensor network. Explicitly,
\begin{equation}\label{ent}
S_A \propto c n \propto c\tau= c \log_2 \tilde N,
\end{equation}
which is precisely the logarithm law in one dimensional system, and $c$ is the proportional constant which is related to the logarithm of the dimensions of the Hilbert space of every link. The cut line (dashed line in Fig.\ref{fig:2}) can be viewed as the reminiscence of the Ryu-Takayanagi(RT) surface in AdS/CFT \cite{RT}.

We have seen that deep neural network has the power to encode holography. There are some basic reasons why we need deep neural network but not shallow neural network (SNN) which has only one layer of hidden neurons:

(i) DNN can efficiently represent most physical state, while shallow neural network is unable\cite{GD};

(ii) When the system is at critical point, the number of hidden neurons grows obviously to obtain accurate approximation of the physical state \cite{TM}. This qualitatively agrees with RG scheme. And quantum field theory at critical point can always be described by conformal field theory from which AdS can emerge;

(iii) Any (local) tensor network state has a (local) neural network representation if the number of hidden neurons is sufficiently large\cite{HM}. And one special tensor network which can encode holography is multi-scale entanglement renormalization ansatz (MERA) \cite{er,swingle}.
 
 Based on these observations, we conclude that, to encode holography, we need DNN, instead of SNN.

\section{Discussion}
In this essay, we show that spacetime in some sense can be viewed as an emergence from deep neural network of quantum state and the entanglement structure is of the RT form. This gives rise to an alternative way towards understanding the basic mechanism of the holography.

There are some interesting future directions to study. For example, the deep neural representation of the partition function of the boundary quantum field theory
\begin{equation}
\mathcal Z= \sum_{\{v_i\},\{h_j\}}e^{-\mathbf E(\{v_i\},\{h_j\})}=\sum_{\{v_i\},\{h_j\}}e^{-\sum_j b_j h_j - \sum_{ij} v_i w_{ij} h_j -\sum_i c_i v_i}.
\end{equation}
may also be recognized as bulk gravitational partition function. Particularly, $\{h_j\}$ can be regarded as a bulk field which is an external source of the boundary CFT field $\{v_i\}$. The term $\sum_{ij} v_i w_{ij} h_j$ may be understood as $\int \mathrm d^d x \phi(x) O(x)$ term in GKP-W dictionary \cite{gkp,witten}, where $\phi(x)$ is bulk field and $O(x)$ is boundary field. This is the power of deep neural network and up to now tensor network does not have access to GKP-W dictionary which plays a key role in AdS/CFT.

Another interesting future problem is the causal structure of deep neural network. Just like MERA \cite{beny}, deep neural network may exhibit intrinsic causal structure and give rise to Lorentz signature, then the time axis may emerge from \emph{depth} of deep neural network.

\acknowledgments

We are very grateful to Hong Guo and Chong-Bin Chen for stimulating discussions during the preparation of the essay. This work was supported in part by the National Natural Science Foundation of China under Grant No. 11465012.

\end{document}